\documentstyle[prb,aps,psfig]{revtex}
\begin{document}
\draft
\twocolumn[\hsize\textwidth\columnwidth\hsize\csname
@twocolumnfalse\endcsname

\title{Spatially homogeneous ground state of the two-dimensional Hubbard 
model}
\author{Federico Becca, Massimo Capone, and Sandro Sorella}
\address{
Istituto Nazionale per la Fisica della Materia and 
International School for Advanced Studies, Via Beirut 4,
34013 Trieste, Italy\\}

\date{\today}
\maketitle
\begin{abstract}
We investigate the stability with respect to phase separation or charge
density-wave formation of
the two-dimensional Hubbard model for various values of the local Coulomb
repulsion and electron densities using Green-function Monte
Carlo techniques.
The well known sign problem is particularly serious in 
the relevant region of small hole doping. 
We show that the difference in accuracy for different doping makes
it very difficult to probe the phase separation instability using only
energy calculations, even in the weak-coupling limit ($U=4t$) 
where reliable results are available.
By contrast, the knowledge of the charge correlation functions allows us to
provide clear evidence of a spatially homogeneous ground state up to $U=10t$.
\end{abstract}
]

\section{Introduction}

Since the discovery of high-temperature superconductivity, the two-dimensional
Hubbard model \cite{hubbard} has been the subject of a
huge amount of work. Indeed, it is widely accepted that, despite the 
complicate structure of these materials, a key role is played by the 
electron correlation in the ${\rm CuO}$ planes, well represented by 
the Hubbard model.

This model and the related $t-J$ model reproduce, at least
qualitatively, some of the physical properties of cuprates. For 
example, at half-filling (i.e., when the number of electrons is equal to the
number of sites) the ground state (GS) is an antiferromagnetic insulator,
and upon doping the antiferromagnetism is strongly suppressed. 
The possible insurgence of superconductivity upon doping and the 
symmetry of the order parameter in the model are still 
open questions.
Another important point to study is the stability with respect to 
phase separation (PS) or charge-density waves (CDW) near half-filling. 
Indeed, both PS and CDW have been experimentally observed in different 
systems \cite{jorgensen,tranquada} and different authors 
\cite{emery,castellani} have pointed out
a possible relation between charge instability and superconductivity.

Because of the strongly interacting nature of these systems, important
insights came from nonperturbative numerical methods.

The exact diagonalization methods, e.g., Lanczos, are strongly limited 
by the exponential 
growth of the Hilbert space, and in practice it is possible to diagonalize 
only up to about 20 sites. 
A remarkable development of the exact diagonalization strategy is the density
matrix renormalization group \cite{white}, which allows us to compute the
GS using an iteratively improved basis.
Despite the accuracy of this technique for one-dimensional
and quasi-one-dimensional systems, there is no straightforward 
generalization for higher dimensions.

Attractive alternatives are stochastic methods such as
Quantum Monte Carlo (QMC) that do not suffer from severe limitations
in terms of lattice size and hence allow us to study fairly large systems.
The simplest QMC simulation is the variational one, in which it is 
possible to compute the expectation values of the energy and 
all the correlation functions for a given wave function. 
The limitation of this method is that one has to guess {\it a priori}
the form of the GS.

QMC methods allow us to sample directly the GS by using projection
techniques, the simplest one being the so-called power method, namely
the simple idea that an iterative application of the Hamiltonian $H$
filters out the ground state starting from an arbitrary state
nonorthogonal to it. These approaches suffer, in fermionic systems,
from the so-called sign problem. Indeed, due to the
antisymmetry of the wave function under permutations of two particles,
one gets after a few power iterations opposite contributions leading
to the cancellation of large and fluctuating weights. 

An alternative approach, which is quite efficient for the Hubbard model,
is the auxiliary field quantum Monte Carlo (AFQMC) \cite{afqmc}.
The Coulomb interaction is linearized by using a Hubbard-Stratonovic
transformation. In this way it is possible to apply exactly the
operator $e^{-\beta H}$ to a Slater determinant.
In practice, AFQMC is very accurate only for small values
of the Coulomb repulsion, whereas, when the interaction 
strength becomes comparable with
the bare bandwidth, large fluctuations make the simulations rather
unstable and inefficient, because the sign problem becomes 
particularly severe for strong Coulomb repulsion.

A sophisticated implementation of the power method 
is the Green-function Monte Carlo (GFMC) applied to lattice models
\cite{nandini}. 
In principle, this method gives unbiased results but in
practice, for fermion systems, the sign problem makes 
any simulation prohibitive, because of statistical errors. 
Recently, many attempts have been made to deal with fermions
and overcome the sign problem \cite{cpqmc,fn}. 
The fixed-node approximation \cite{fn} 
is introduced for the  GFMC and replaces the original Hamiltonian
with an effective one free of the 
sign problem, in which the nodes are fixed to be those 
of the so-called guiding wave function.
In practice, the most common choice is the best variational
wave function available, i.e., the one with the lowest energy. 
The fixed-node energy is an upper bound to the true GS energy and 
the method is variational.
Moreover, it becomes exact if the guiding function is
chosen to be the exact GS function. Although this 
approximation can be uncontrolled and needs an external input, 
it is quite accurate with respect to both energy and correlation functions 
for small sizes. 

An alternative approach, 
recently proposed by one of us, is to use the GFMC with a stochastic
reconfiguration (SR) \cite{sorella}.
The main idea of this method is to use a suitable reference dynamics 
free of the sign problem to constrain the true dynamics.  
The fixed-node and a variational dynamics are natural 
candidates as references.
At each reconfiguration the amplitudes of the GS
are chosen as small perturbations of the reference wave function
amplitudes such that the mixed averages of a few physically
relevant quantities are conserved. The method is exact if
all the operators in the Hilbert space are reconfigured, but it has been
shown that a major role is played by a very small number of 
operators \cite{sorcap}.

In this paper,
we present a systematic study on the Hubbard model comparing different 
QMC techniques by studying in particular the stability of PS 
and CDW near half-filling
for various values of the Coulomb repulsion.
In Sec. \ref{model} we introduce the Hubbard model and we discuss
the implementation of our approximations, in Sec. 
\ref{results} we report our results,
and finally in Sec. \ref{conclusions} we give a brief summary.

\section{The Model}\label{model}

We consider the Hubbard model on a square lattice of $L$ sites with 
$N=N_{\uparrow}+N_{\downarrow}$ particles and  $N_{\uparrow}=N_{\downarrow}$,
where $N_{\uparrow}$ ($N_{\downarrow}$) is the number of spin-up (-down) particles.
In order to study PS as close as possible to half-filling 
using only closed-shell configurations, we consider
square lattices tilted by $45^\circ$ with $L^2=2 l^2$ and $l$ odd.
In this way half-filling is a closed shell and the first doped closed shell
has eight holes independent of $L$.
The Hamiltonian reads
\begin{equation}\label{hamilt}
H = -t \sum_{\langle i,j \rangle, \sigma} c^{\dag}_{i,\sigma} c_{j,\sigma}
+ U \sum_{i} n_{i,\uparrow} n_{i,\downarrow},
\end{equation}
where $\langle \; \rangle$ stands for nearest neighbors, $c_{i,\sigma}$ 
($c^{\dag}_{i,\sigma}$) destroys (creates) an electron with spin $\sigma$ at
site $i$, and $n_{i,\sigma}=c^{\dag}_{i,\sigma}c_{i,\sigma}$. 
In the  following, all energies are measured in units of $t$. 

As already pointed out above, in the presence of the sign problem 
the choice of the guiding wave function is crucial.
Our wave function reads
\begin{equation}\label{afwf}
|\Psi_{G} \rangle = {\cal P}_{Sz=0} {\cal P}_g {\cal J} |{\cal D} \rangle ,
\end{equation}
where $|{\cal D} \rangle$ is a Slater determinant in which the orbitals are suitably 
chosen (see below), ${\cal P}_{Sz=0}$ is the projector onto the
subspace with $N_{\uparrow}=N_{\downarrow}$, i.e., with zero total spin
component along the $z$ axis,
${\cal P}_g$ is a Gutzwiller operator that inhibits 
the double occupancies ${\cal P}_g = exp(-g\sum_{i} n_{i\uparrow}n_{i\downarrow})$,
where $g$ is a variational parameter, and $\cal J$ is a Jastrow factor
${\cal J} = exp(\frac{\gamma}{2}\sum_{i,j} v_{i,j} S_i^z S_j^z)$, 
where $\gamma$ is another variational parameter and $v_{i,j}$ is taken from 
spin-waves theory \cite{franjic}.
Care must be taken in the choice of the orbitals appearing in 
the Slater determinant.
The most common choice is to take
the orbitals from a Hartree-Fock (HF) approximation of the Hamiltonian
breaking the SU(2) spin rotation symmetry along the $z$ axis. 
\begin{equation}\label{hf}
H = -t \sum_{\langle i,j \rangle, \sigma} c^{\dag}_{i,\sigma} c_{j,\sigma}
+ \frac{U}{2} \sum_{i, \sigma} 
\left [ 
\langle n_i \rangle - \sigma (-1)^{R_i} \langle m_i \rangle
\right ] n_{i,\sigma},
\end{equation}
where
\begin{eqnarray}
\langle n_i \rangle & = & \langle n_{i,\uparrow} \rangle +
\langle n_{i,\downarrow} \rangle, \\
\langle m_i \rangle & = & (-1)^{R_i} \left [
\langle n_{i,\uparrow} \rangle -
\langle n_{i,\downarrow} \rangle \right ].
\end{eqnarray}
We consider only fillings which are closed shells for $U=0$ 
and where a solution with constant density 
$\langle n_i \rangle=\frac{N}{L}=n$ and staggered magnetization 
$\langle m_i \rangle=m$ are found.
In this case, the HF many-body wave function can be written as
\begin{equation}\label{det1}
|{\cal D} \rangle = 
\prod^{1,...,N_{\uparrow}}_k \beta^{\dag}_{k,\uparrow}
\prod^{1,...,N_{\downarrow}}_q \beta^{\dag}_{q,\downarrow} |0 \rangle,
\end{equation}
where the quasiparticles have definite momentum modulo $Q=(\pi,\pi)$ and
definite spin, since 
the antiferromagnetic order parameter is along the $z$ axis
\begin{equation}
\beta^{\dag}_{k,\sigma}=
u_k c^{\dag}_{k,\sigma}+\sigma v_k c^{\dag}_{k+Q,\sigma},
\end{equation}
$k$ is in the reduced magnetic Brillouin zone, and
$u_k$ and $v_k$ are defined in Ref.~\cite{shiba}. It is worth noting that
for $U/t \rightarrow \infty$, $u_k=v_k=\frac{1}{\sqrt{2}}$, namely the
spin up and the spin down are in different sub-lattices 
(classical N\'eel state).

In a previous work \cite{cosentini} the wave function 
(\ref{afwf}) with $|{\cal D} \rangle$ given by Eq.~(\ref{det1}) has been 
found to be a rather poor approximation for large $U/t$ at half-filling.
In particular, in this representation the Jastrow factor ${\cal J}$ does
not play any important role.

We propose a wave function which is a straightforward generalization
of the one successfully used for the Heisenberg model \cite{calandra}.
The fundamental ingredient is to allow spin 
fluctuations perpendicular to the staggered magnetization. An easy 
implementation of this idea is to put the magnetization 
in the $x-y$ plane allowing transverse fluctuations along the $z$ axis 
through a Jastrow-like factor \cite{franjic}.
This is achieved by a $\frac{\pi}{2}$ rotation $U_y(\frac{\pi}{2})$ 
around the $y$ axis of
the canonical operators:
\begin{eqnarray}
U^{\dag}_y(\frac{\pi}{2}) c^{\dag}_{i,\uparrow} U_y(\frac{\pi}{2}) & = & 
\frac{1}{\sqrt{2}} 
\left ( c^{\dag}_{i,\uparrow} + c^{\dag}_{i,\downarrow} \right ), \\
U^{\dag}_y(\frac{\pi}{2}) c^{\dag}_{i,\downarrow} U_y(\frac{\pi}{2}) & = & 
\frac{1}{\sqrt{2}} 
\left ( c^{\dag}_{i,\uparrow} - c^{\dag}_{i,\downarrow} \right ).
\end{eqnarray}
The fermionic part of our guiding wave function is therefore defined 
as a Slater determinant of the transformed orbitals,
\begin{eqnarray}
\label{rotated}
\beta^{\dag}_{k,+} &=& 
U^{\dag}_y(\frac{\pi}{2}) \beta^{\dag}_{k,\uparrow} U_y(\frac{\pi}{2}), \\
\label{rotated2}
\beta^{\dag}_{k,-} &=&
U^{\dag}_y(\frac{\pi}{2}) \beta^{\dag}_{k,\downarrow} U_y(\frac{\pi}{2}),
\end{eqnarray}
namely it is given by
\begin{equation}\label{det2}
|{\cal D} \rangle = 
\prod^{1,...,N_{\uparrow}}_k \beta^{\dag}_{k,+}
\prod^{1,...,N_{\downarrow}}_q \beta^{\dag}_{q,-} |0 \rangle.
\end{equation}
Remarkably for $U/t \rightarrow \infty$ and half-filling, by construction 
Eq.~(\ref{det2}) becomes the N\'eel state with spin
quantization parallel to the $x$ axis, i.e. it has the correct
Marshall sign on each of the $2^L$  configurations sampled by GFMC.
In this limit, it is also clear why the Jastrow factor may be much more 
effective: being defined along the $z$ axis, 
it allows us to sample the quantum fluctuation perpendicular to 
the staggered magnetization. In the previous case instead both  the Jastrow 
quantization axis and the order parameter were 
parallel, and for $U/t \rightarrow \infty$ there is no way to sample any
fluctuation, the only possible configuration being  the classical one.

It is worth noting that, although for the $t-J$ model the d-wave BCS
wave function with a spin-rotationally invariant density-density 
Jastrow factor represents
a very accurate variational state, for the Hubbard model at small and 
intermediate coupling ($U \le 10t$) the best choice for the variational and
guiding wave function is given by a Jastrow-Slater determinant with rotated
orbitals Eqs.~(\ref{rotated}) and (\ref{rotated2}). Indeed, although the d-wave
BCS wave function is a singlet and does not break the SU(2) symmetry, it has
a very poor variational energy for the Hubbard model. 
The quality of the variational energy obtained with our 
Jastrow-Slater determinant remains considerably better than the 
BCS one at half-filling and $U \le 10t$ even when the accuracy 
of the approximation is improved by the GFMC.
Instead, in the doped case, the BCS wave function with GFMC 
is only slightly worse than the corresponding Jastrow-Slater determinant 
proposed in this work. This may suggest that antiferromagnetism is already 
suppressed at small finite doping, and d-wave superconductivity is a 
possible stable phase especially at large $U/t$. 

An important systematic improvement of the wave function can be achieved
by performing exactly one Lanczos step starting from $|\Psi_G \rangle$,
\begin{equation}
\label{LS}
|\Psi_L \rangle = \left ( 1+\alpha H \right ) |\Psi_G \rangle,
\end{equation}
with $\alpha$ free parameter chosen to minimize the energy.
This technique has been successfully used for the $t-J$ model both to improve
the variational calculation \cite{heeb} and as a starting point for power
methods \cite{chen}. Henceforth, we will denote by VMC and LS the results 
obtained with the wave function Eqs.~(\ref{afwf}) and (\ref{LS}), respectively, 
using variational Monte Carlo. Analogously, the symbols FN and FNLS will
indicate the fixed-node approximation applied to the wave function 
Eqs.~(\ref{afwf}) and (\ref{LS}), respectively.
Finally, the symbols SR will denote the stochastic reconfiguration approximation
applied to the wave function Eq.~(\ref{LS}).

In Tables \ref{tabella} and \ref{tabella2} 
we report the energies of 18 and 10 electrons on 18 sites, 
respectively. At half-filling we compare the results using Eq.~(\ref{afwf})
with the Slater determinant $|{\cal D} \rangle$ given by Eqs.~(\ref{det1}) 
and (\ref{det2}) for different
approximations and values of $U/t$. Using Eq.~(\ref{det2}) we obtain 
a sizeable improvement for large $U$'s ($U \ge 10t$). Notice that for $U=20t$
the best variational result (the FNLS) with Eq.~(\ref{det1}) is worse than the 
simple VMC with Eq.~(\ref{det2}). For 10 electrons the two Slater determinants
give the same results. Indeed, for this doping the antiferromagnetic order
is strongly suppressed and the Jastrow factor ${\cal J}$ does not play any
important role.

\section{Results}\label{results}

One of the most debated issues in strongly correlated electron models
is the nature of the charge distribution in their GS.
Recently, many authors \cite{lee,lin,putikka,hellberg,kohno,cbs,rommer} 
have addressed the question of PS in the $t-J$ model.
It is well accepted that for $J \gg t$, holes tend to cluster together
leaving the rest of the system in an antiferromagnetic state 
without holes. Although most of the calculations lead to the
conclusion that a critical value of $J$ below which
the GS is homogeneous exists, it is not clear what this value is at low
doping, ranging
from $0.5t$ and $1.2t$. Moreover, some authors \cite{cdw} have suggested
that just before PS, the GS has charge modulations.
Recently, an accurate numerical study of a few chains on the Hubbard
model \cite{bonca} has shown evidence of stripes, i.e., CDW oscillations.
However, this result appears limited to quasi-one-dimensional
geometry, as also suggested by the authors.

It is well known that PS is characterized by an infinite compressibility
in the thermodynamic limit. 
The compressibility can be related to the curvature of the energy
with respect to the electron density,
\begin{equation}\label{energy}
\chi = \left ( {\partial^2 E \over \partial n^2} \right )^{-1}.
\end{equation}
From the above definition we have 
that a divergent $\chi$ corresponds to a vanishing curvature of the 
energy as a function of density.
Therefore, PS can in principle be detected by means of energy measurements
for various densities. This is an appealing property,
since the accuracy on the energy is usually better than that of any other 
observable for most numerical methods.
Many previous numerical studies of PS have therefore concentrated
on the energy curve, or equivalently, on the energy per hole
$e_h(\delta)= [e(\delta) - e_H]/\delta$, 
where $e(\delta)$ is the energy per site at a hole density $\delta=1-n$ and
$e_H=e(0)$ is the energy at half-filling \cite{lin}. 
If $\chi$ diverges, $e_h(\delta)$ is flat in the thermodynamic limit
and develops a minimum for $\delta=\delta_c$ in finite systems, due
to the finite positive surface energy at the phase boundary.

This approach has been pursued by Cosentini {\it et al.} \cite{cosentini},
using FN calculations. They found that there is a large region of PS 
in the phase diagram, at least for
$U \ge 10t$, clearly in contrast with what is found in the $t-J$ model.
In this paper, we consider the Hubbard model and we show that a study
of PS instability is very difficult using only energy calculation.
Instead, a careful calculation of charge-correlation functions 
strongly indicates that the GS is homogeneous.

In order to show that the energy calculations may overestimate the
tendency to a PS instability, it is important to compare the GFMC 
results with some exact reference results.
Previous studies have shown 
that it is important to consider relatively large lattice
sizes since finite-size effects favor PS \cite{imada}.  
We need, therefore, a reference result for large lattices, 
where exact diagonalization is not available. 
In the case of the Hubbard model for small $U$, the AFQMC is almost 
exact and represents the reference we need.
As stated in the preceding section, we consider only closed shell doping 
because, at least for small $U$, huge finite size 
effects affect the physical properties in a rather drastic way.
For instance, the large bare density of states near half-filling 
determines an unphysical and spurious PS up to the first closed
shell \cite{imada}.

\begin{figure}
\centerline{\psfig{bbllx=50pt,bblly=250pt,bburx=500pt,bbury=650pt,%
figure=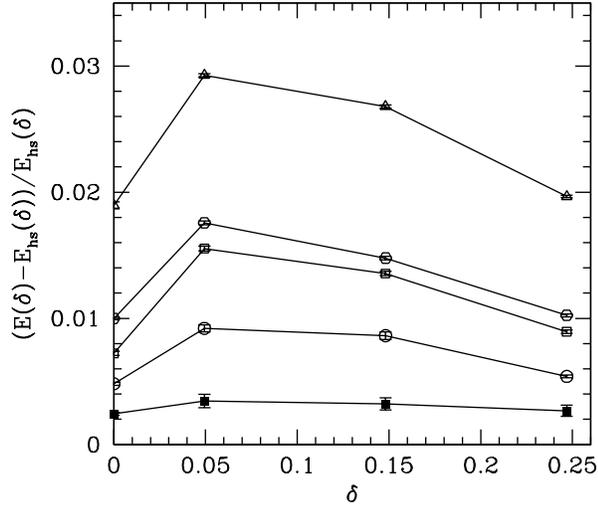,width=80mm,angle=0}}
\caption{\baselineskip .185in \label{accuracy}
Relative accuracy of various GFMC techniques with respect to AFQMC
for a 162-site lattice with $U=4t$ as a function of filling $\delta$.
From top to bottom VMC (empty triangles), 
FN (empty hexagons), LS (empty squares),
FNLS (empty circles), and SR (full squares). Lines are guides to the eye.}
\end{figure}

In Fig.~\ref{accuracy}, we show the accuracy of the GFMC results obtained 
with different approximations
compared with the AFQMC ones for a 162-site lattice
and $U=4t$. 
For this coupling value, AFQMC does not provide evidence for PS.
We plot $[E(\delta)-E_{hs}(\delta)]/E_{hs}(\delta)$,
where $E(\delta)$ and $E_{hs}(\delta)$ are the energies of GFMC and AFQMC,
respectively, for a doping $\delta$.
Besides the improvement in the absolute accuracy, 
the curves get flatter and flatter improving the
approximation, but only the SR accuracy is almost doping-independent. 
In other words, we need a very accurate calculation to 
eliminate the spurious dependence of the 
variational energy upon doping \cite{note}.
Even for the best variational method, the FNLS, although the 
accuracy on the energy is for all doping less than $1\%$, 
the difference in accuracy between, for example, the half-filled case
and the first closed shell is still sizable.
This difference is very important, because it represents just the energy scale
determining or ruling out PS.

\begin{figure}
\centerline{\psfig{bbllx=50pt,bblly=250pt,bburx=500pt,bbury=650pt,%
figure=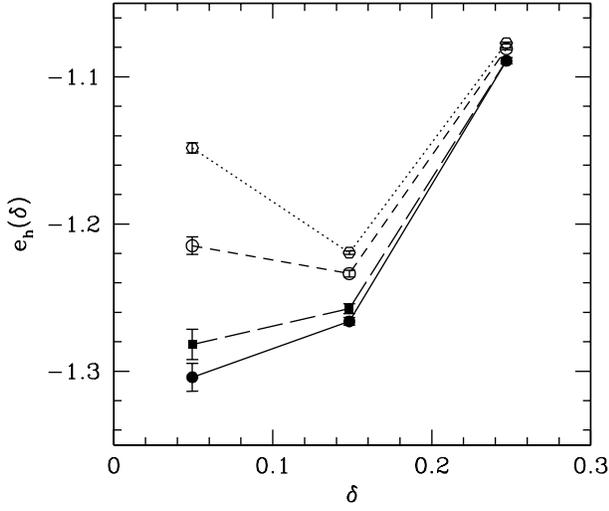,width=80mm,angle=0}}
\caption{\baselineskip .185in \label{emery}
Energy per hole $e_h(\delta)$ for a 162-site lattice with $U=4t$.
From top to bottom FN (empty hexagons), FNLS (empty circles), 
SR (full squares) , and AFQMC (full circles). Lines are guides to the eye.}
\end{figure}

In Fig.~\ref{emery}, the function $e_h(\delta)$ is shown for the
FN, FNLS, SR, and AFQMC methods.
We need to use SR to exclude the occurrence of PS, 
where even the FNLS data would imply PS. 
The reason for this disappointing situation is that all 
the known variational methods are still
too dependent on the guiding wave function. 
With the previous analysis, the resolution in energy necessary to detect 
or rule out PS is very hard
to reach with statistical methods, especially for large $U/t$.

On the other hand, GFMC methods have proven to be reliable
not only for energy calculations but also for correlation functions such as 
$N(q) = \langle n_q n_{-q}\rangle$,
where $n_q$ is the Fourier transform of the electron density \cite{cbs}.
For a phase-separated system, there are strong fluctuations in the density for
small momenta and $N(q \rightarrow 0)$ is expected to be strongly
enhanced for small momenta, that is, for $|q| \sim \frac{2\pi}{\xi}$, where
$\xi$ is the characteristic length of the phase-separated region.
Moreover, if $\chi$ diverges, also $N(q \rightarrow 0)$ diverges, yielding
an alternative tool to probe PS.

\begin{figure}
\centerline{\psfig{bbllx=50pt,bblly=250pt,bburx=500pt,bbury=650pt,%
figure=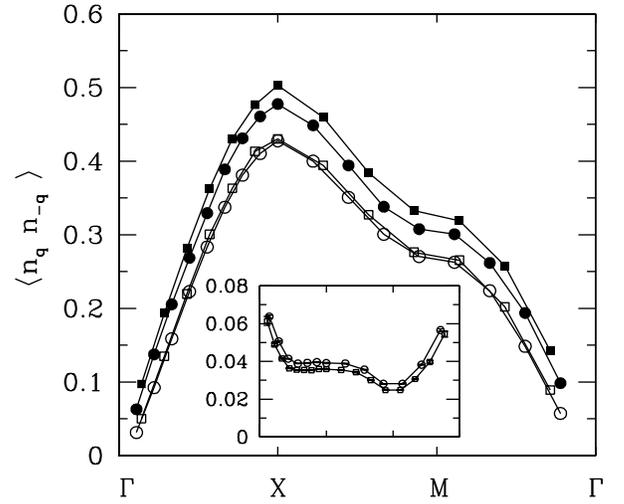,width=80mm,angle=0}}
\caption{\baselineskip .185in \label{nqu4}
$N(q)$ for $U=4t$, 162 electrons on 162 sites (empty circles), 98 electron
on 98 sites (empty squares), 154 electrons on 162 sites (full circles)
and 90 electrons on 98 sites (full squares). Lines are guides to the eye and
error bars are smaller than points.  $\Gamma = (0,0)$, $X = (\pi,\pi)$, $M = (\pi,0)$.
In the inset: $N(q)$ for the $t-J$ model, $J=0.6t$, 156 electrons on 162 sites 
(squares), and 94 electrons on 98 sites (circles).}
\end{figure}

This method turns out to be more reliable, since it is 
based on a single calculation for a given doping value, whereas the
evaluation using $e_h(\delta)$ involves a comparison between energies obtained
by different simulations for different fillings, with the corresponding
guiding wave function having different accuracies.
Indeed with GFMC, $N(q)$ has been
proved to be a very sensitive tool to look at for detecting PS. 
In the $t-J$ model, $N(q)$ has a very different shape 
for stable and unstable systems. 
A clear peak at the smallest $q$ indicates PS even when, as shown in the
inset of Fig.~\ref{nqu4}, $J/t$ is very close to the PS boundary \cite{cbs}. 
Moreover, from $N(q)$ it is also possible to extract information about
charge fluctuations at finite $q$'s, related to CDW. 
In Ref.~\cite{cbs} we have
shown that in the $t-J$ model for $J=0.4t$, $N(q)$ has some peaks at
finite $q$'s, surprisingly near to what was found in recent experiments 
\cite{peak}. The knowledge of $N(q)$ allows us to extract more
general results with respect to the simple study of $e_h(\delta)$.
Furthermore, $N(q)$ is found to be much less size dependent than $e_h(\delta)$.
For the $t-J$ model there is no appreciable difference between 98-site
and 162-site lattices and for $J=0.4t$. 
Both calculations suggest that there is no
PS, whereas an analysis using the FN approximation of $e_h(\delta)$ should
lead to PS for 98 sites and to a homogeneous state for 162 sites.
Indeed, in this case PS is only a size effect and a homogeneous state
is found by increasing the accuracy of the method or 
by increasing the lattice size.

We computed $N(q)$ by means of the forward-walking technique \cite{calandra},
within the FNLS approximation, at half-filling
and for the first few closed-shell configurations on a 162- and a 
98-site lattice. 
The evaluation of the density-density correlation function is in principle possible
even within SR by numerical differentiation of the energy with respect to an
external field coupled to $N(q)$. However, this approach is very
demanding and does not give a significant improvement on the FNLS 
results, which are very accurate. Indeed for the smallest $q$ vector 
for 90 electrons on the
98-site lattice, i.e., $q=(2\pi /7,2\pi /7)$, we found the the AFQMC gives
$N(q)=0.0932(2)$ and the FNLS gives $N(q)=0.096(1)$. 
 
In Fig.~\ref{nqu4}, $N(q)$ is shown for $U=4t$ at half-filling for
a 162-site and a 98-site lattice
and for 154 electrons on a 162-site lattice and 90 electrons on 
a 98-site lattice.
No sign of divergence, and consequently of PS or CDW, is seen in the data.
We also notice that the two sets of points for the half-filled systems 
lie on the same curve, showing that we have 
substantially reached the thermodynamic limit.
For this value of $U$, $N(q)$ is essentially featureless for all 
doping we considered, suggesting that there are no charge
instabilities at any finite length. The smallest doping we considered
is $\delta \simeq 0.049$ and we cannot exclude that for smaller
doping PS or CDW are present. 

\begin{figure}
\centerline{\psfig{bbllx=50pt,bblly=250pt,bburx=500pt,bbury=650pt,%
figure=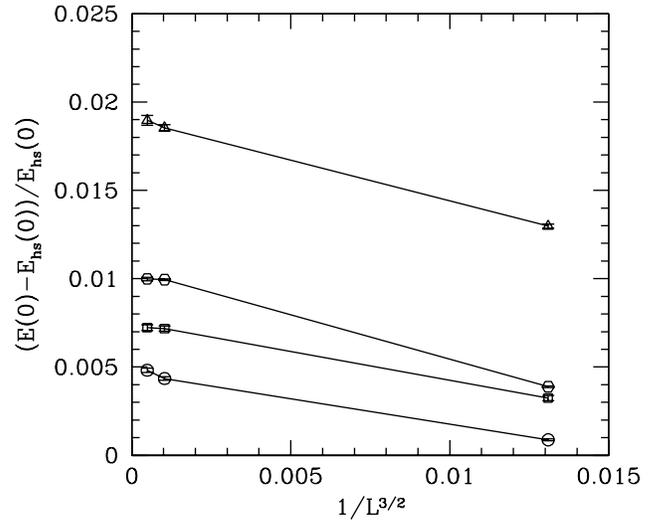,width=80mm,angle=0}}
\caption{\baselineskip .185in \label{halffilling}
Relative accuracy of various GFMC techniques with respect to AFQMC
for different lattices ($L=18,98,162$) and $U=4t$.
From top to bottom: VMC (empty triangles), 
FN (empty hexagons), LS (empty squares),
and FNLS (empty circles). Lines are guides to the eye.}
\end{figure}

In order to investigate smaller doping,
we should consider larger lattices. Unfortunately, the accuracy of the
approximations considered decreases when increasing the size of the system
and the 162-site lattice represents
the largest lattice where the accuracy is acceptable. 
In Fig.~\ref{halffilling},
we report $[E(0)-E_{hs}(0)]/E_{hs}(0)$ for various sizes and for different
approximations: from the 18 sites to the 162 sites, the accuracy 
of FNLS changes from less than $0.1\%$ to about $0.5\%$. These indications
prevent us from considering sizes larger than the ones presented in this
paper.

Now we turn to larger Coulomb interactions and consider $U=10t$, 
where the AFQMC results are not reliable due to large
fluctuations.
In principle, GFMC techniques do not suffer from intrinsic limitations in the
large-coupling regime and it is possible to consider any value of $U$.
In practice we need an accurate knowledge of the nodes, i.e.,
an accurate guiding wave function.
Our choice, Eq.~(\ref{afwf}), with orbitals given by Eqs.~(\ref{rotated}) and
(\ref{rotated2}) is a very good approximation for the half-filling
case. In Table \ref{tabella}, we report the energies for various methods
for 18 electrons on 18 sites at different $U$'s. Although all the
approximations are quite size-dependent, the wave function becomes more
and more accurate by increasing the Coulomb potential.
Therefore, we expect that it also gives a good starting point at least
close to half-filling.

We present results for $U=10t$, for which previous FN calculations based on
$e_h(\delta)$ and a less accurate wave function \cite{cosentini} have
shown PS up to $\delta\simeq 0.15$. Indeed, if we use 
$e_h(\delta)$ as a probe
for PS, we find that the phase diagram shows a large instability region,
confirming the results of Ref.~\cite{cosentini}.
As for the $U=4t$ case, this instability is very likely to be a spurious
effect, a consequence of the different energy accuracy for different doping. 
This possibility, which cannot be proved without knowing the exact
energies at strong coupling (at present impossible), is instead
very clearly supported by the calculation of the charge-correlation
functions.

\begin{figure}
\centerline{\psfig{bbllx=50pt,bblly=250pt,bburx=500pt,bbury=650pt,%
figure=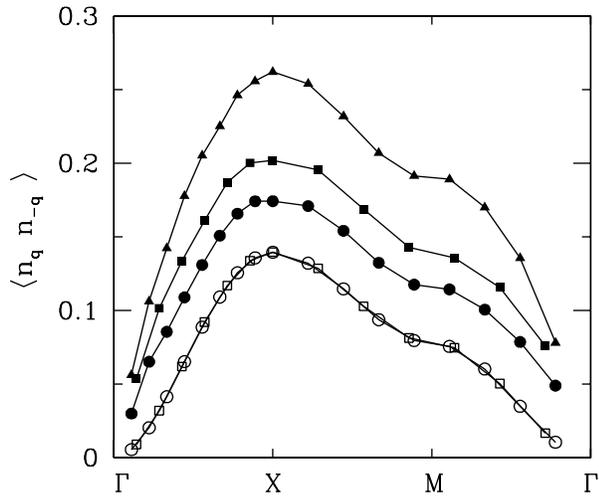,width=80mm,angle=0}}
\caption{\baselineskip .185in \label{nqu10}
$N(q)$ for $U=10t$, 162 electrons on 162 sites (empty circles), 98 electrons
on 98 sites (empty squares), 154 electrons on 162 sites (full circles),
90 electrons on 98 sites (full squares), and 138 electrons on 162 sites 
(full triangles). Lines are guides to the eye and error bars are smaller
than points.
$\Gamma = (0,0)$, $X = (\pi,\pi)$, $M = (\pi,0)$.}
\end{figure}

Figure \ref{nqu10} displays $N(q)$ for the same fillings of Fig.~\ref{nqu4}
and for 138 electrons on the 162-site lattice,
which corresponds to $\delta\simeq 0.148$.
All the correlation functions are definitely nondivergent for 
$q \rightarrow 0$ and are qualitatively similar to the $U=4t$ case,
indicating that the system is far away from a PS instability.
Furthermore $N(q)$ does not show peaks at any finite momenta 
for this Coulomb interaction.
This finding shows that the Hubbard and the $t-J$ model may have 
different behaviors as far as charge correlations are concerned.
Indeed, by diagonalizing exactly the 18-site lattice, we find that charge
fluctuations for the Hubbard and for the $t-J$ model are quite different
in the small doping region for $U=10t$ and $J=0.4t$, respectively \cite{becca}.

\section{Conclusions}\label{conclusions}

An extensive GFMC analysis of the Hubbard model at low hole doping
has been carried out. In particular, we have focused on the possible
instability of the model with respect to PS and CDW.
Comparing GFMC results with AFQMC ones in the weak-coupling region 
($U=4t$), we show that detecting PS by means of
energy results requires a very accurate calculation at all electronic
densities.
Indeed, the accuracy of the energy is strongly
dependent on the electron density, and 
the signature of PS based only on energy calculations 
is clearly affected by this bias, 
leading to a spurious region of PS instability.
In the case of the Hubbard model, this is particularly relevant
because, while we are able to give a very good description of the
half-filled case, in which the GS is an antiferromagnetic insulator,
we are not aware of equally accurate descriptions of the doped state.
Even for $U=4t$, it is necessary to use the really accurate SR 
and AFQMC technique
to eliminate the doping dependence of the accuracy and to rule out PS.

On the other hand, PS (and CDW) instability can be probed more easily
using charge correlation functions. This approach has various advantages.
First, it is found that $N(q)$ has very small size effects and the
thermodynamic limit is reached with about 100 sites, both for the Hubbard
and the $t-J$ model. Second, the information contained in 
$N(q)$ does not depend on different densities, implying that a different
accuracy as a function of doping does not introduce any external bias.

Instead of using energy calculations, which are very expensive at moderate
and large $U$'s, we calculate the charge correlation functions and we
are able to find clear evidence for the absence of PS up to
$U=10t$ in the low doping regime.

\acknowledgments

It is a pleasure to acknowledge useful discussions with L. Capriotti, 
M. Calandra, G. Bachelet, E. Koch, and A. Parola, to whom we are also 
grateful for critical and careful reading of the manuscript.
This work has been partially supported by MURST-COFIN99 and by
Istituto Nazionale per la Fisica della Materia.

\begin{table}
\begin{tabular}{lllllll}
$U$   & $|{\cal D} \rangle $ & $E_{ex}$ & $E_{VMC}$  & $E_{FN}$   & $E_{LS}$   & $E_{FNLS}$  \\
\hline \hline
4t    &  (\ref{det1})        & -0.9585 & -0.9382(1) & -0.9514(1) & -0.9520(1) & -0.9556(1) \\
10t   &  (\ref{det1})        & -0.4484 & -0.4034(1) & -0.4284(1) & -0.4154(1) & -0.4316(1) \\
20t   &  (\ref{det1})        & -0.2339 & -0.2023(1) & -0.2195(1) & -0.2060(1) & -0.2225(1) \\
4t    &  (\ref{det2})        & -0.9585 & -0.9460(1) & -0.9547(1) & -0.9553(1) & -0.9576(1) \\
10t   &  (\ref{det2})        & -0.4484 & -0.4382(1) & -0.4451(1) & -0.4428(1) & -0.4470(1) \\
20t   &  (\ref{det2})        & -0.2339 & -0.2293(1) & -0.2232(1) & -0.2310(1) & -0.2337(1) 
\end{tabular}
\caption{GS energies for 18 electrons on 18 sites as a function of $U/t$ using (\ref{det1})
and (\ref{det2}) as Slater determinant.}
\label{tabella}
\end{table}

\begin{table}
\begin{tabular}{llllll}
$U$   & $E_{ex}$ & $E_{VMC}$  & $E_{FN}$   & $E_{LS}$   & $E_{FNLS}$  \\
\hline \hline
4t    & -1.1299 & -1.1124(1) & -1.1218(1) & -1.1229(1) & -1.1263(1) \\
10t   & -1.0193 & -0.9749(1) & -1.0006(1) & -0.9997(1) & -1.0098(1) \\
20t   & -0.9598 & -0.8983(1) & -0.9354(1) & -0.9253(1) & -0.9450(1)
\end{tabular}
\caption{GS energies for 10 electrons on 18 sites as a function of $U/t$.}
\label{tabella2}
\end{table}



\begin{thebibliography}{99}

\bibitem{hubbard} J. Hubbard, Proc. R. Soc. London, Ser. A {\bf 276}, 
   238 (1963); M.C. Gutzwiller, \prl {\bf 10}, 159 (1963); 
   J. Kanamori, Prog. Theor. Phys. {\bf 30}, 275 (1963).
\bibitem{jorgensen} J.D. Jorgensen, B. Dabrowski, Shiyon Pei, D.G. Hinks,
   L. Soderholm, B. Morosin, J.E. Shirber, E.L. Venturini, and 
   D.S. Ginley, \prb {\bf 38}, 11337 (1988).
\bibitem{tranquada} J.M. Tranquada, B.J. Sternlib, J.D. Axe, Y. Nakamura,
   and S. Uchida, Nature {\bf 375}, 561 (1995).
\bibitem{emery} V.J. Emery and S.A. Kivelson, Physica C {\bf 209}, 
   597 (1993).
\bibitem{castellani} C. Castellani, C. Di Castro, and M. Grilli, \prl {\bf 75},
   4650 (1995).
\bibitem{white} S.R. White, \prl {\bf 69}, 2863 (1992); S.R. White, \prb
   {\bf 48}, 10345 (1993).
\bibitem{afqmc} D.R. Hamann and S.B. Fahy, \prb {\bf 41}, 11352 (1990);
   S.B. Fahy and D.R. Hamann, \prl {\bf 65}, 3437 (1990).
\bibitem{nandini} N. Trivedi and D.M. Ceperley, \prb {\bf 41}, 4552 (1990).
\bibitem{cpqmc} Shiwei Zhang, J. Carlson, and J.E. Gubernatis, \prl {\bf 74},
   3652 (1995); Shiwei Zhang, J. Carlson, and J.E. Gubernatis, \prb {\bf 55},
   7464 (1997).
\bibitem{fn} D.F.B. ten Haaf, H.J.M. van Bemmel, J.M.J. van Leeuwen,
   W. van Saarloos, and D.M. Ceperley, \prb {\bf 51}, 13039 (1995).
\bibitem{sorella} S. Sorella, \prl {\bf 80}, 4558 (1998).
\bibitem{sorcap} S. Sorella and L. Capriotti, \prb {\bf 61}, 2599 (2000).
\bibitem{franjic} F. Franjic and S. Sorella, Prog. Theor. Phys. {\bf 97},
   399 (1997).
\bibitem{shiba} H. Yokoyama and H. Shiba, J. of Phys. Soc. of Japan. 
   {\bf 56}, 3582 (1987).
\bibitem{cosentini} A. Cosentini, M. Capone, L. Guidoni, and G. Bachelet,
   \prb {\bf 58}, 14685 (1998).
\bibitem{calandra} M. Calandra and S. Sorella, \prb {\bf 57}, 11446 (1998).
\bibitem{heeb} E.S. Heeb and T.M. Rice, Europhys. Lett. {\bf 27}, 673 (1994).
\bibitem{chen} Y.C. Chen and T.K. Lee, \prb {\bf 51}, 6723 (1995).
\bibitem{lee} C.T. Shih, Y.C. Chen, and T.K. Lee, \prb {\bf 57}, 627 (1998).
\bibitem{lin} V.J. Emery, S.A. Kivelson, and H.Q. Lin, \prl {\bf 64},
   475 (1990).
\bibitem{putikka} W.O. Putikka, M.U. Lucchini, and T.M. Rice, \prl {\bf 68},
   538 (1992).
\bibitem{hellberg} C.S. Hellberg and E. Manousakis, \prl {\bf 78}, 
   4609 (1997).
\bibitem{kohno} M. Kohno, \prb {\bf 55}, 1435 (1997).
\bibitem{cbs} M. Calandra, F. Becca, and S. Sorella, \prl {\bf 81}, 5185 
   (1998).
\bibitem{rommer} S. Rommer, S.R. White, and D.J. Scalapino, \prb {\bf 61}, 13424 (2000).
\bibitem{cdw} S.R. White and D.J. Scalapino, \prl {\bf 80}, 1272 (1998);
   {\bf 81}, 3227 (1998).
\bibitem{bonca} J. Bonca, J.E. Gubernatis, M. Guerrero, 
   E. Jeckelmann, and S.R. White, \prb {\bf 61}, 3251 (2000).
\bibitem{imada} N. Furukawa and M. Imada, J. of Phys. Soc. of Japan
   {\bf 61}, 3331 (1992).
\bibitem{note} We checked that without rotating orbitals, that is,
   using the wave function of Ref.~\cite{cosentini}, 
   we have much less accuracy as a function of density.
\bibitem{peak} J.M. Tranquada, J.D. Axe, N. Ichikawa, A.R. Moodenbaugh,
   Y. Nakamura, and S. Ucida, \prl {\bf 78}, 338 (1997).
\bibitem{becca} F. Becca, A. Parola, and S. Sorella, \prb {\bf 61}, 16287 (2000).
\end{thebibliography}
\end{document}